\documentstyle{article}
\input amssym.def
\input amssym

\begin{document}

\centerline{\Large \bf
ON THE CLASSICAL LIMIT}

\centerline{\Large \bf OF THE BALANCED STATE SUM}

\bigskip
\centerline{\parbox{56mm}{Louis Crane and David N. Yetter \\ Department of
Mathematics \\ Kansas State University \\ Manhattan, KS 66506}
\footnote{Supported by NSF Grant \# DMS-9504423}  }
\bigskip

\noindent{\bf I. Introduction}

\bigskip

In [1], a new state sum  model (the ``balanced'' model) was suggested
as a quantum theory of gravity. In [2], a proposal was made to
construct a geometrical interpretation of it, and to demonstrate that
it gives general relativity in the classical limit. 

The purpose of this note is to make several advances in the
interpretation of the balanced model. First, we outline a shortcoming
of the definition of the model in [1] pointed out to us by John
Barrett and John Baez [3], and explain how to correct it. Second, we
show that the classical limit of our state sum reproduces the
Einstein-Hilbert lagrangian whenever the term in the state sum to
which it is applied has a geometrical interpretation. Next we outline
a program to demonstrate that the classical limit of the state sum is
in fact dominated by terms with geometrical meaning. This uses in an
essential way the alteration we have made to the model in order to
fix the shortcoming discussed in the first section. Finally, we make a
brief discussion of the Minkowski signature version of the model.

This note is not intended to be self contained, it can only be read
after familiarizing oneself with [1] and [2].

\bigskip

\noindent{\bf II. Fixing a Hole}

\bigskip

As pointed out in [3], the argument leading to the definition of the
balanced model overlooked a geometric fact. Specifically, if we have
four bivectors $b_i$ ; i= 1,2,3,4, which satisfy the constraints
proposed in [1], namely

\[ |b_{i+}| = |b_{i-}| \]

\[ | (b_i+b_j)_+|= |(b_i+b_j)_-| \]

\[ b_1+b_2+b_3+b_4 =0, \]

\noindent it does not follow that they are the four bivectors corresponding to
the faces of a tetrahedron in ${\Bbb R}^4$ as stated in [1]. Another
configuration is also possible, in which the planes associated to the
4 bivectors all intersect in a common line.

In the classical theory it is easy enough to eliminate this other set
of possibilities. An elegant way to do so is to form the trivector
whose components are the square roots of the four determinants defined in [2]. In the
desired case, when the four bivectors are associated to the faces of a
tetrahedron, we obtain the volume trivector of the tetrahedron by this
formula. In the other case, which we wish to avoid, the components of
the trivector are all zero. Thus, in order to eliminate the unwanted
cases, we project out the intersection of the null spaces of the
operators given by the four determinants of [2].
This also eliminates tetrahedra of
volume zero, which we also do not want in our theory.

This prescription has a natural extension to the quantum theory, since
the definitions of the four determinants are unplagued by ordering
problems.

Thus, to our list of constraints above we add, on each tetrahedron,
the condition $|F|>0$ , where F is the trivector of the (quantum )
tetrahedron.

Of course, it remains to be seen by a careful analysis exactly what
quantum states this leaves us.

\bigskip

\noindent{\bf III. The Classical Lagrangian}

\bigskip

A critical step in the argument of [1] is motivated only by
heuristic arguments, and by the extreme naturality of the proposal.
Namely, after replacing the bivectors on the faces of a 4-simplex with
pairs of spins and replacing the classical constraints with operators
in the category of representations of ${\frak so}(4)$, we close the maps so
defined into a spin net, called a balanced 15j symbol, which we
evaluate to obtain a number. A state sum is now formed as a sum of
products of  these numbers, which we propose as a quantum version of general
relativity.

Our immediate purpose here is to show that this proposal does in fact
correspond to a discrete approximation to the Einstein-Hilbert action
in a suitable classical limit. Our argument closely parallels the 3d
analysis of [4].

More precisely, let us imagine that we have a term in our state sum,
i.e. an assignment of representations to the faces and tetrahedra of a
triangulated 4-manifold whose absolute values closely approximate the
numbers calculated from the areas and dihedral angles of a classical
state of the regge calculus for the triangulation, or otherwise put,
of a choice of flat metrics on each 4-simplex which match at all the
tetrahedra where they intersect. Let us call such a term a ``geometric
configuration''. We want to show that under a small modification of
the data, the evaluation of the balanced 15j symbol varies in a way
which closely approximates the variation of the Einstein-Hilbert action.

Our point of departure in demonstrating this is the well-known
asymptotic formula for the evaluation of a 6j symbol for ordinary
representations of ${\frak su}(2)$:

\[  6j \sim  \frac{1}{\sqrt{12 \pi V}} cos( \Sigma J_i \theta
            _i + \pi /4 ) ,\]

\noindent where we have used the values of the casimir on the six irreducible
representations as lengths for a euclidean tetrahedron, and the $
\theta _i $ and V are the dihedral angles and volume. This
approximation is in a picture where the couplings of the spins are
around faces rather than at vertices, as is more usual; thus it is a
topologically dual picture. This is important in what follows. All of this is
corresponding to the classically permitted region for a 6j symbol.

The dependence of this expression on the cosine of the dihedral angle
is what allowed Regge and Ponzano to construct 3d quantum gravity from
a state sum of spin nets in [4]. The key point is that for the state sum built
of 6j symbols the variation
as a single spin (quantum edge
length) is varied vanishes if the sum of the dihedral angles around
the edge is  $2 \pi $.

What we would really like to show for the balanced state sum is that,
as we vary as single edge length, the variation of the phase of a term 
corresponding to a geometric
configuration is a discretized approximation to the Ricci
curvature of the 4d discete metric of the
configuration. Unfortunately, that is quite difficult to show. The
reason is that there is no quantity in the balanced state sum which
directly represents an edge length. As we shall discuss below, there
is a natural edge length operator, but we have not yet investigated
it.

Thus, we are going to show only that for a certain family of
deformations of a geometric configuration, the change in the 15j
symbol is the same as the corresponding change in the Einstein-Hilbert
lagrangian. Since the family of deformations is sufficient to generate
all allowable changes in the term, this suggests strongly that the effective
lagrangian of the theory is a discrete version of E-H. The
demonstration that we in fact have a quantum theory whose classical
solutions correspond to discretized general relativity (i.e. the
recovery of Einstein's equations) will still
require considerable work. We give a program for demonstrating that
below.

Accordingly, then, let us assume that we have a 4-simplex with
balanced spin labels on its faces and tetrahedra, which correspond to
the geometric data on a euclidean 4-simplex. Now let us consider
a certain type of perturbation. The specific type of perturbation we
consider is to increase or decrease all three of the spins on the
faces incident to a single edge by 1. If we choose to divide the
tetrahedra incident to the line in such a way that the three faces are
on the same side of each division, then the move we are making
corresponds to adding a spin which circulates around a simple three
edged loop. (We can always choose to devide each tetrahedron any way
we like, because the sum of projections on the balanced summands in
any two subdivisions are equal.) It is 
known from the theory of the binor calculus that
every spin net can be decomposed as a superposition of such circulating
spins. Hence our family of deformations, combined with judicious
redivisions of the tetravalent vertices on tetrahedra, can generate
all deformations of a 15j symbol. Let us call such a perturbation a
``basic'' perturbation.

Now let us use the decomposition theory of $\frak so(4)$ spin networks to
rewrite the balanced 15j symbol as a product of a 6j and a 12j symbol,
where the 6j unites the spins on the 3 faces surrounding an edge with
the 3 bivectors representing their sums. (See Figure 1.)

This has the effect of
isolating all the spins to be changed in our simple perturbation onto
a single 6j symbol. Thus, the asymptotic formula for the evaluation of
a 6j symbol (or more precisely its square, since we are considering
representations of ${\frak so}(4) \cong {\frak su}(2)\oplus {\frak su}(2)$) 
allows us to approximate the effect of our deformation.

Let us now imagine a euclidean tetrahedron whose edge lengths are dual
to the 
spins on the 6j symbol we have isolated; so that the edges
corresponding to the three faces meet at a vertex instead of forming a
triangle. This dualization allows us to use the above formula to study
the behavior of the
associated 6j symbol under perturbation. Let us call this the
``ancillary tetrahedron.'' The effect of our deformation now can be
approximated very simply as the sum of the three areas to be deformed
each times the derivative of the cosine of the corresponding dihedral
angle for the ancillary tetrahedron.

Now the crucial step in the analysis is the geometrical interpretation
of the dihedral angles of the ancillary tetrahedron. On the one hand,
the three planar angles connecting the edges corresponding to the
three faces are approximately the angles between the three classical
bivectors corresponding to the three faces. On the other, they form a
spherical triangle with the three dihedral angles of the ancillary
tetrahedron.

Thus, the three dihedral angles of the ancillary tetrahedron are
determined from the three angles between the three classical bivectors
by the laws of spherical trigonometry.

On the other hand, if we think of the geometry of the classical
4-simplex, the angles between the three bivectors on the faces around
an edge have the same relationship with the three hyperdihedral angles
on those faces: they are the edge angles and spherical angles of a
spherical triangle. To see this, we pick any point interior to the edge, and
intersect the 4-simplex with the hyperplane orthogonal to the edge at
the point. The intersection with the three faces and tetrahedra around
the edge form three planes, whose intersection with a small sphere
centered at the point is a spherical triangle. (See Figure 2)

Thus, since the dihedral angles of the 4-simplex are closely
approximated by the face angles of the ancillary tetrahedron, it
follows that the dihedral angles of the ancillary triangle are good
approximations to the hyperdihedral angles of the 4-simplex. Thus the
variation formula for the 6j symbol of the ancillary tetrahedron shows
us that the effect of one of our basic perturbations is the sum of the
areas of the faces times the derivatives of the cosines of the
hyperdihedral angles on them.

If we now consider a geometric configuration for the state sum on a
whole triangulated 4-manifold, the result of the forgoing analysis is
that the basic deformations have the same effect as the change in the
Einstein-Hilbert lagrangian.

Let us try to see why this is so. The Einstein-Hilbert lagrangian for
a 4 dimensional manifold is the sum of the sectional curvatures of the
manifold, with the measure given by the metric. When one considers the
setting of a discretized metric, the curvature is distributional, and
concentrated on the 2-simplices of the triangulation. Thus the Regge
version of the discretized E-H action is the sum of the areas of the
2-simplices times the curvature of a section transverse to them, which
is just the deficit angle, ie the sum of the hyperdihedral angles
surrounding the face minus $2 \pi$. The terms $2 \pi A$ can be
neglected in a discretized setting, since the areas are integral. We
can then rewrite the discretized EH lagrangian for a triangulated
4-manifold as a sum of expressions for each 4-simplex, where the
expression for each 4-simplex is just the sum of the area of each
2-simplex bounding the 4-simplex, times the corresponding
hyperdihedral angle. 

Now by a well known theorem in classical geometry [4], the variation in
this expression as we vary one area is just the corresponding
hyperdihedral angle. Thus the variation in the contribution to our
lagrangian from one 4-simplex when we make our basic perturbation is
just the sum of the three hyperdihedral angles incident on the edge we
have chosen.

To relate this to the result using the 6j symbol on the ancillary
tetrahedron, it is necessary to recall that the 15j symbol is not a
contribution to the lagrangian, but to a path integral, hence to the
complex exponential of the lagrangian. Thus we are interested not in
the variation of the 6j symbol itself, but of its phase, $ \Sigma J_i
\theta_i$.

The analogous classical theorem in 3d now gives us that its variation
is also $ \theta _i$ when we vary a single J, or the sum of the three
angles for the basic deformation.

This argument is a strong indication, at this relatively early stage,
that the balanced state sum is really a quantum version of general
relativity. From the standpoint of proof (even as physicists use the
term), the argument is not very powerful. What is needed is an
argument that geometric configurations are plentiful, and in fact
produce all the classical solutions.

Now we sketch a program for repairing these shortcomings.

\bigskip

\noindent{\bf IV. Towards a quantum geometry}

\bigskip

Since we have included in our revision of the balanced state sum the
constraint that the trivector operators on the tetrahedra are nonzero,
we can form the components of four of the trivectors into a $4 \times
4$ matrix and compute its determinant. This gives us an operator for
the hypervolume of the 4-simplex. The next thing we would like to
investigate would be the inverse of this matrix; this would generate
operators corresponding to edge lengths. At this point operator
ordering problems will have to be faced in the quantum theory.

The task of showing that we have many geometrical configurations in
the classical limit boils down to a study of the operator algebra of
the operators corresponding to all the classical geometrical
quantities on a 4-simplex: lengths areas, dihedral angles, volumes and
hyperdihedral angles and hypervolume. If the commutators between these
operators can be neglected, then their simultaneous eigenvectors will
provide geometrical configurations. The observation in [5] that the
commutators of the operators determining dihedral angles vanish
because of cancellations in balanced tetrahedra make us
optimistic. Much work will be needed in this direction.

\bigskip

\noindent{\bf V. The Minkowski signature case}

\bigskip

The reduction of the deformations of 15j symbols to 6j symbols above
suggests that the extension of the euclidean model above to a
lorentzian one will parallel the situation in 3d. There, the
classically forbidden regime of 6j symbols models the geometry of
lorentzian tetrahedra much the same way that classically allowed 6j
symbols model euclidean ones. This allowed Barrett and Foxon to
construct 3d lorentzian general relativity as a categorical state sum [6].

It seems highly likely that a similar procedure will work in 4d as
well. The state sum will be over the same labelling set, but a wick
rotation of the evaluations will give the dominant weight to
lorentzian rather than euclidean configurations. 

The first step in the analysis will be relating the lornetzian geometry
of the 4-simplex with that of the ancillary tetrahedron.

\bigskip

\centerline{\bf Bibliography}
\bigskip

1. J.W. Barrett and L. Crane, ``Relativistic Spin Networks and Quantum
Gravity,'' gr-qc 9709028 (to appear CQG).

\smallskip

2. L. Crane, ``On the Geometric Interpretation of Relativistic Spin
Networks,'' gr-qc 9710108.

\smallskip

3. private communication.

\smallskip

4. G. Ponzano and T. Regge, ``Semiclassical Limit of Racah
Coefficients'' in {\em Spectroscopic and Group Theoretic Methods in Physics}
(R. Bloch, ed), North Holland, Amsterdam, 1968.

\smallskip

5. A. Barbieri, ``Quantum Tetrahedra and Simplicial Spin Networks,'' gr-qc
9707010.

\smallskip

6. J.W. Barrett and T.J. Foxon, ``Semiclassical Limits of Simplicial
Quantum Gravity,'' gr-qc 9310016.

\smallskip

7. L. Crane, L.H. Kauffman and D.N. Yetter, ``State Sum Invariants of
4-Manifolds,'' {\em JKTR} vol. 4 No. 2 (1997) 177-234.

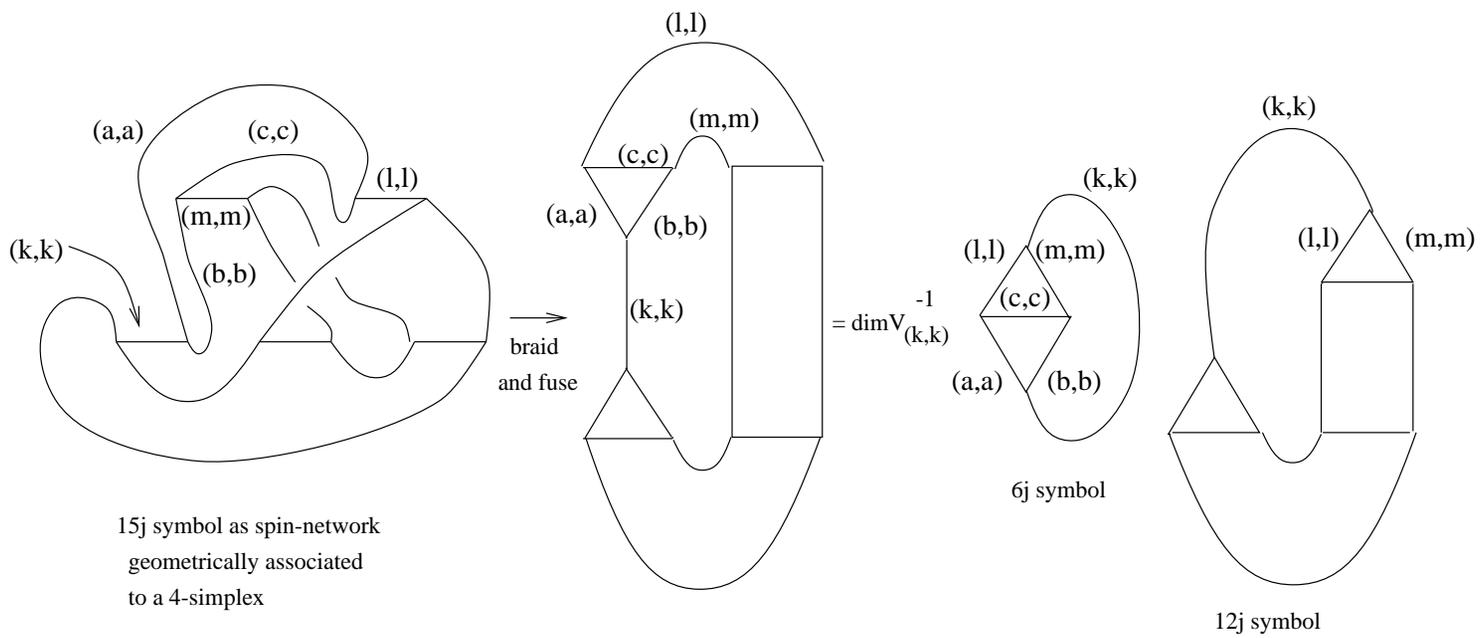
\end{document}